# Extremely High Thermal Conductivity of Aligned Polyacetylene Predicted using First-Principles-Informed United-Atom Force Field

Teng Zhang,[1,2,*] Jiaxin Xu,[1] Tengfei Luo[1,3,*]

1. Department of Aerospace and Mechanical Engineering, University of Notre Dame, Notre Dame, Indiana 46556, United States
2. Schrödinger Inc., New York, NY 10036 USA
3. Department of Chemical and Biomolecular Engineering, University of Notre Dame, Notre Dame, Indiana 46556, United States

* Corresponding author: T. Zhang: zhteg4@gmail.com; T. Luo: tluo@nd.edu

**Abstract:**
Molecular simulations of polymer rely on accurate force fields to describe the inter-atomic interactions. In this work, we use first-principles density functional theory (DFT) calculations to parameterize a united-atom force field for polyacetylene (PA) – a conjugated polymer potentially of high thermal conductivity. Different electron correlation functionals in DFT have been tested. Bonding interactions for the alternating single and double bonds in the conjugated polymer backbone are explicitly described and Class II anharmonic functions are separately parameterized. Bond angle and dihedral interactions are also anharmonic and parameterized against DFT energy surfaces. The established force field is then used in molecular dynamics (MD) simulations to calculate the thermal conductivity of single PA chains and PA crystals with different simulation domain lengths. It is found that the thermal conductivity values of both PA single chains and crystals are very high and are length-dependent. At 790 nm, their respective thermal conductivity values are ~480 W/mK and ~320 W/mK, which are comparatively higher than those of polyethylene (PE) – the most thermally conductive polymer fiber measured up-to-date.

## 1. Introduction

The discovery of polyacetylene (PA) has led to the launch of the field of organic conductive polymers.[1] PA is a representative of the Π-conjugated polymers. Π-conjugated polymers have a number of outstanding chemical and physical properties, such as high electrical conductivity and high modulus. They are widely used in many areas, including chemical sensors, light-emitting diode, photovoltaics and cancer treatment.[2-7] Recent experiments have found that ultra-drawn polyethylene (PE) nanofibers have thermal conductivity three orders of magnitude higher than its bulk amorphous counterpart due to highly aligned and straight chain morphologies.[8, 9] More recent experiments showed that other polymers fibers, especially Π-conjugated polymers, can have thermal conductivities even higher than PE.[10] Simulations were able to qualitatively explain the chemistry-thermal conductivity relation by linking the molecular structures to thermal conductivity.[11-13] The stiff backbone of Π-conjugated polymers, especially the dihedral angles, makes the chains much less susceptible to segmental rotation and thus allows phonons to travel for a long distance besides the high phonon group velocities enabled by the strong bonds.[14] In the practical perspective, the ability to tailor the thermal transport properties of polymers is critical in a wide range of applications, such as thermal management of electronics and energy conversion.[15, 16]

The next step in this field is to more quantitatively predict the upper limit of thermal conductivity of polymers so as to offer valuable guidance for selecting and synthesizing materials to achieve high thermal conductivity. This relies on accurate atomistic potential models. The COMPASS potential (condensed-phase optimized molecular potentials for atomistic simulation studies)[17] has been previously used to simulate many polymers (*e.g.* PE) in condense phase or isolated chains, and the development of this force flied is mainly based on *ab-initio* calculations of small molecules.[18] Unlike PE, the binding properties of Π-conjugated polymers, such as bond strength, angle bending, and backbone rigidity, can be a strong function of degree of polymerization due to the delocalized electrons over the whole molecule. Such dependence on length questions the accuracy of COMPASS potential for Π-conjugated polymer simulation. In the long chain limit, a key feature of the Π-conjugated polymer is the alternating single and double bonds. The COMPASS potential, however, does not distinguish the types of these bonds. Other force fields like CHARMM[19] and AMBER[20] are not suitable for thermal transport study, since they consider bond and angle with only harmonic terms. As a result, a more accurate potential with higher order terms is desired for the thermal transport study in Π-conjugated polymers.

A force field specifically developed for PA can provide more reliable estimations of the backbone rigidity and bond strength in Π-conjugated polymers, and the roles of Π-conjugated structures in thermal transport can be better evaluated. In this work, we use *ab-initio* calculations to produce highly accurate benchmark data against which a united-atom force field is parameterized. Compared to the COMPASS potential, this potential can more accurately describe the bonding nature of the PA and is capable of distinguishing correctly the single and double bonds. Using such a potential, we predict the thermal conductivity of single PA chains and crystals. At 800 nm, their respective thermal conductivity values are ~500 W/mK and ~320 W/mK, which are comparatively higher than those of PE – the most thermally conductive polymer fiber measured up-to-date. We also found dramatically different values compared to those from COMPASS potentials.

## 2. Computational Method

All *ab-initio* calculations were carried out using the GAMESS (2012) software,[21, 22] except that for the infinitely long chain limit calculations, the Gaussian software[23] which has periodic boundary condition implemented was used. Six different basis sets and three different electronic correlation methods were tested. Becke's three-parameter exchange functional combined with the LYP correlation functional (B3LYP) and the 6-31G(d) basis set at the restricted Hartree-Fock (RHF) self-consistent field (SCF) level of theory were found to give accurate results with affordable computational cost and were used for producing benchmark data for PA.[24-26] A modified united atom Class II potential model[27] was parameterized against the benchmark data which is capable of describing the special alternating single and double bonding feature. All the molecular dynamics (MD) simulations were carried out using the large-scale atomic/molecular massively parallel simulator (LAMMPS)[28] with a time step of 0.25fs, and non-equilibrium molecular dynamics (NEMD) was used to calculate the thermal conductivity of PA single chains and crystals.

## 3. Results and Discussion

### 3.1 Basis Set and Exchange Correlation Functional Testing

A combination of six basis sets (MINI, 3-21G(d), 6-31G(d), 6-31+G(pd), 6-331++G(pd), and aug-cc-pVTZ) and three different electronic correlation methods (HF, B3LYP, and CCSDT) are tested on the smallest π-conjugated molecule, 1, 3-Butadiene (Figure 1a) to identify the most efficient yet accurate combination. Efficiency is considered since in the following study much larger systems are studied. Geometry optimizations are first performed using different combinations. The gradient convergence tolerance for optimization is set to be $10^{-4}$ Hartree, and the density matrix convergence threshold is set to be $10^{-5}$ Hartree for all calculations. Total energies, bond lengths and computational efficiency using different combinations are then compared (Table 1).

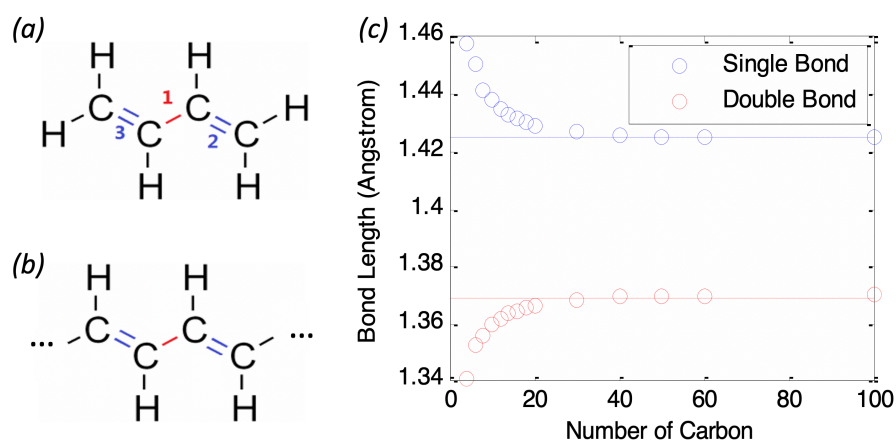

*Figure 1. (a) Chemical formula of 1, 3-Butadiene. The single backbone bond is lighted in red, and the double bond length is obtained by averaging bond 2 and bond 3. (b) Chemical formula of PA. (c) ab-initio calculated bond lengths for single and double bonds in PA with different chain lengths.*

Table 1. Total energy, bond length and computational time consumption of 1, 3-Butadiene calculated using different electronic correlation methods and basis sets.

| Method | Basis | $E_{total}$ (Hartree) | Single Bond (Angstrom) | Double Bond (Angstrom) | Time Used (second) |
|---|---|---|---|---|---|
| HF | MINI | -153.8570468 | 1.505706 | 1.337760 | 2.5 |
|  | 3-21G(d) | -154.0594602 | 1.467108 | 1.320432 | 3.2 |

|       | 6-31G(d)      | -154.9196462 | 1.467638 | 1.322548 | 17.6     |
|-------|---------------|--------------|----------|----------|----------|
|       | 6-31+G(pd)    | -154.9306017 | 1.467257 | 1.322020 | 41.1     |
|       | 6-331++G(pd)  | -154.9623713 | 1.467563 | 1.323270 | 121.3    |
|       | aug-cc-pVTZ   | -154.9801898 | 1.464214 | 1.319759 | 5748.2   |
| B3LYP | MINI          | -154.8630704 | 1.500101 | 1.367187 | 142.2    |
|       | 3-21G(d)      | -155.0308404 | 1.461928 | 1.339167 | 156.5    |
|       | **6-31G(d)**  | **-155.88246** | **1.457750** | **1.340820** | **305.5** |
|       | 6-31+G(pd)    | -155.8929595 | 1.457332 | 1.340057 | 354.8    |
|       | 6-331++G(pd)  | -155.9331869 | 1.456649 | 1.338492 | 582.6    |
|       | aug-cc-pVTZ   | -155.9478729 | 1.453000 | 1.334661 | 9070     |
| CCSDT | MINI          | -154.1570722 | 1.526838 | 1.389285 | 199.6    |
|       | 3-21G(d)      | -154.469657  | 1.481370 | 1.354614 | 3709.3   |
|       | 6-31G(d)      | -155.4872408 | 1.464782 | 1.349376 | 17374.5  |
|       | 6-31+G(pd)    | -155.5390988 | 1.463725 | 1.348560 | 67902.2  |
|       | 6-331++G(pd)  | -155.6006974 | 1.467560 | 1.351192 | 306953.1 |
| Experimental value[17] | | | 1.476 | 1.337 | |

From the above comparison, B3LYP and the 6-31G(d) basis set are found to give sufficiently accurate results within affordable computational cost, and thus are chosen for the following calculations.

## 3.2 Length Effect of Polyacetylene in ab-initio Calculation

For possible thermal transport applications, long chains in the highly crystalline fiber form are needed. In order to approximate the long chain limit, H-(CH)$_n$-H chains with $n$ ranging from 4 to 100 are studied. The lengths of the C-C and C=C in the middle of the chains are showed in Figure 1b, and the dash lines are the results from the calculation using periodic boundary conditions in Gaussian.[13] It is found that the middle bond lengths converge when $n$ is above 40. Therefore, it is safe to derive the force field based on the structures in the middle portion of H-(CH)$_{40}$-H chains. It is worth noting that in the long chain limit, the molecular weights are very high and the majority of the segments will not feel end effects. Parameterizing a potential for the middle bonds will be applicable to simulating such long chains. From Figure 1c, it is also very clear that the single and double bonds have distinct bond lengths, indicating the need of distinguishing them in the MD simulations.

## 3.3 Ab-initio Energy Hypersurface and Force Field Parameterization

To create an energy hypersurface (energy-coordinate relation) for force field parameterization, bond length, angle and dihedral angle of the C=C-C=C segment in the middle portion of H-(CH)$_{40}$-H (red, Figure 2a) are changed and corresponding energies are calculated. (Figure 2b, c, d).

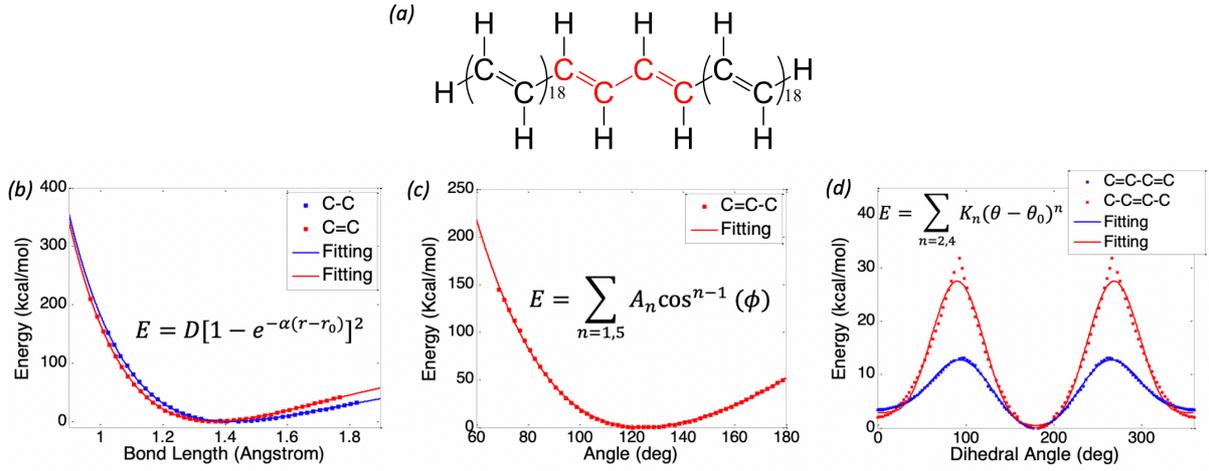

***Figure 2. (a)*** *Force field for polyacetylene is developed based on the middle segment of a H-(CH)$_{40}$-H chain (highlighted in red).* ***(b)*** *Bond,* ***(c)*** *angle, and* ***(d)*** *dihedral angle energy hypersurface and fitted potential of PA.*

If hydrogen atoms are considered, there should be 4 types of bonds, 3 types of angle terms, and 6 types of dihedral angle types. However, explicit hydrogen approach increases the complexity of the model. In addition, for thermal transport studies, the high frequency vibrations associate with hydrogen atoms are largely un-excited at room temperatures according to the Bose-Einstein distribution. The united-atom model lumps hydrogen atoms into the carbon atoms they bind to and thus significantly simplify the MD simulations without introducing significant errors in thermal conductivity prediction. Another reason of ignoring hydrogen atoms is that the hydrogen vibrations are at very high frequencies which are related to optical phonons. Since the group velocities of these optical phonon branches are much smaller than those of the acoustic braches, they usually do not contribute much to the thermal conductivity.[29, 30] Therefore, united atom force field is used, and the above energy hypersurfaces are fitted using empirical functions (Figure 2).

For bonding terms, the Morse potential is used (Equation 1 in Table 2). Class II angle formula (Equation 2 in Table 2) is used to describe the angle interaction. The energy hypersurface is well-fitted by these potential functions (Figure 2b & 2c). Multi-harmonic dihedral style of four *cosine* functions (Equation 3 in Table 2) is found to fit the dihedral angle energy surface well (Figure 2d).

**Table 2.** Parameters of the established united-atom force field for PA.

| Bond | $E = D[1 - e^{-\alpha(r-r_0)}]^2$ (1) where $r_0$ is the equilibrium bond distance, $\alpha$ is a stiffness parameter, and $D$ is the energy constant. | | | |
|---|---|---|---|---|
| | $D$ (Kcal/mol) | $\alpha$ (1/Å) | $r_0$ (Å) | |
| C-C | 100.025 | 2.032 | 1.421 | |
| C=C | 128.263 | 2.068 | 1.368 | |
| Angle | $E = K_2(\theta - \theta_0)^2 + K_3(\theta - \theta_0)^3 + K_4(\theta - \theta_0)^4$ (2) where $\theta_0$ is the equilibrium angle value, and $K_n$ are the energy constants. | | | |
| | $K_2$(Kcal/mol/radian) | $K_3$(Kcal/mol/radian) | $K_4$(Kcal/mol/radian) | $\theta_0$(deg) |
| C=C-C | 83.83 | -52.26 | 23.31 | 124.5 |
| Dihedral | $E = A_1 + A_2 \cos^2(\phi) + A_3 \cos^3(\phi) + A_4 \cos^4(\phi) + A_5 \cos^5(\phi)$ (3) | | | |

|         | where $A_n$ are energy constants. | | | | |
|---------|---------------------|---------------------|---------------------|---------------------|---------------------|
|         | $A_1(Kcal/mol)$ | $A_2(Kcal/mol)$ | $A_3(Kcal/mol)$ | $A_4(Kcal/mol)$ | $A_5(Kcal/mol)$ |
| C=C-C=C | 12.70 | -1.95 | -15.61 | 3.77 | 4.68 |
| C-C=C-C | 27.48 | -1.32 | -49.16 | 2.54 | 23.37 |

### 3.4 Molecular Dynamics Simulations

Morse bond, Class II angle, and multi-cosine dihedral angle are implemented in LAMMPS with the parameters in Table 2. van der Waals (vdW) interactions are simulated using the 12-6 Lennard-Jones potential with parameters from the Polymer Consistent Force Field (PCFF).[31] The Mulliken charges of carbon and hydrogen atoms are 0.1234 and -0.1234 electron charges,[32] respectively, and thus the united-atom model of the -CH- segment is of zero charge. The mass of -CH- is set to be the sum of the mass of one carbon and one hydrogen atom. One single PA chain is constructed by a given number of -CH- units. For a 400-segment PA chain, the length extends to 49.4 nm after minimization. NEMD is then used to calculate the thermal conductivity along the PA chain length direction (Figure 3a). In an NEMD simulation, a temperature gradient is established by thermostating a heat sink and a heat source region at different temperatures. Except a 10 Å thick slice of fixed atoms at each end, the rest of the structure is simulated using NVE ensemble with two Langevin thermostats at the ends: the heat source is set to be 315 ºC and the heat sink is set to be 285 ºC (Figure 3a). The 10 Å-thick slice of fixed atoms is use to prevent the energy transfer across the periodic boundary in the chain length direction. After steady state is reached, heat flux can be calculated using $J=dQ/dt/S$, where $dQ/dt$ is the average energy change rates of the two Langevin thermostats, and $S$ is the single-chain cross sectional area calculated from the PA crystals. The temperature gradient ($dT/dx$) can be obtained by fitting the linear portion of the temperature profile (Figure 3b), and then the thermal conductivity can be calculated using Fourier's law, $\kappa=-J/(dT/dx)$. Different chain lengths have been studied. We have also calculated the PA crystal thermal conductivity. In these calculation, 48 PA chains are packed in a perfectly crystalline phase[33] and then the system is optimized using NPT ensembles with 1 atmosphere as the cross-sectional pressure. The along-chain direction is first stretched to simulate the drawing process, and then the all three dimensions are relaxed at 1 atmosphere and 300K. Similar NEMD scheme is used to thermal conductivity of the single PA chain whose cross-sectional area is 16.18 Å$^2$ .

The thermal conductivity of a 49.36 nm-long single chain PA using our new force field is 226.03 ± 5.63 W/mK. The uncertainty is calculated as the standard deviation from four independent runs. This value is much higher that the ~65 W/mK for ~50 nm-long PE single chain using COMPASS potentials, showing that PA has a higher thermal conductivity than PE with the same length (Figure 3d). In addition, the COMPASS potentials predict the thermal conductivity of a ~50 nm-long PA single chain to be 124.90 W/mK, and detailed comparison between our united-atom model and the COMPASS model is discussed in Section 3.5.

A further study on length dependent thermal conductivity of single PA chains is carried out. The simulated length ranges from 12.34 nm to 789.68 nm and the corresponding thermal conductivity values are presented in Figure 3d. The result shows that the thermal conductivity of single chain PA keeps increasing from 121.69 W/mK to 480.02 W/mK, as the sample length increases from 12.34 nm to 789.68 nm, though the rate of the increase is gradually decreasing. It is possible that, in PA chains, the intrinsic phonon mean free path along the chain length direction is much larger than our simulated length domain, thus the phonon transport can be ballistic and mainly be scattered at chain ends. As a result, the thermal conductivity gets higher as the phonon mean free path increases resulted from the increasing chain length. However, as the chain length gets longer, phonon-phonon scattering becomes more important than boundary scattering, and thus the increasing rate of thermal conductivity slows down.

Similarly, thermal conductivity of crystalline PA increases as a function of chain length, and converge to a much higher value after 400 nm compared to the crystalline PE (Figure 3d), showing that the PA has an outstanding high thermal conductivity in bulk phase for practical applications. The along-chain phonon scattering due to inter-chain interaction in crystals reduces the phonon mean free path compared to that of the single chains.[34] Although boundary scattering still dominants in shorter chains, the smaller intrinsic phonon mean free path than single chain cases makes the thermal conductivity increase as a function of chain length in crystal milder than single chains.

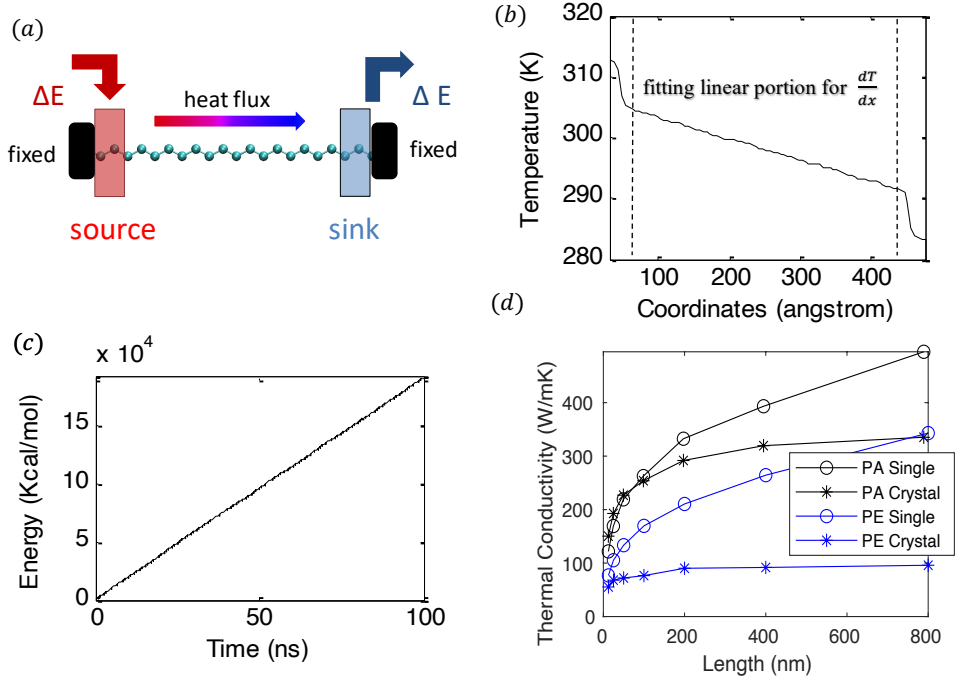

*Figure 3. (a) NEMD scheme, (b) temperature profile, and (c) energy change of Langevin thermostats as a function of time. (d) Thermal conductivity of single chain and crystalline PA and PE with different lengths.*

## 3.5 Phonon Dispersion Relationship Calculation

To understand the root cause of the difference in thermal conductivity predicted from our united-atom model and the COMPASS potential, we calculate the phonon dispersion using these two models. With the minimized structures in section 3.4, the single chains are first simulated at 2 K with fixed volume, and then simulated in ensembles of constant volume and energy (NVE) for 50 ps. During the NVE runs, the velocities of all backbone atom are recorded every 5 fs. For the phonon dispersion relationship calculation, two-dimensional Fourier transform of the atomic velocity of one atom in the unit cell is first performed:

$$\Phi(\omega, k) = \sqrt{\sum_{\alpha}^{3} \left| \frac{1}{N} \sum_{n=0}^{N-1} e^{i\frac{n}{N}k} \int v_\alpha(n,t) e^{-i\omega t} dt \right|^2}, \alpha = x, y, z \quad (4)$$

where $v_\alpha(n,t)$ is the atomic velocity, $\omega$ the frequency, $k$ is the wavevector, $n$ is the index of repeating unit along the chain direction, and $N$ is the number of the repeating unit.

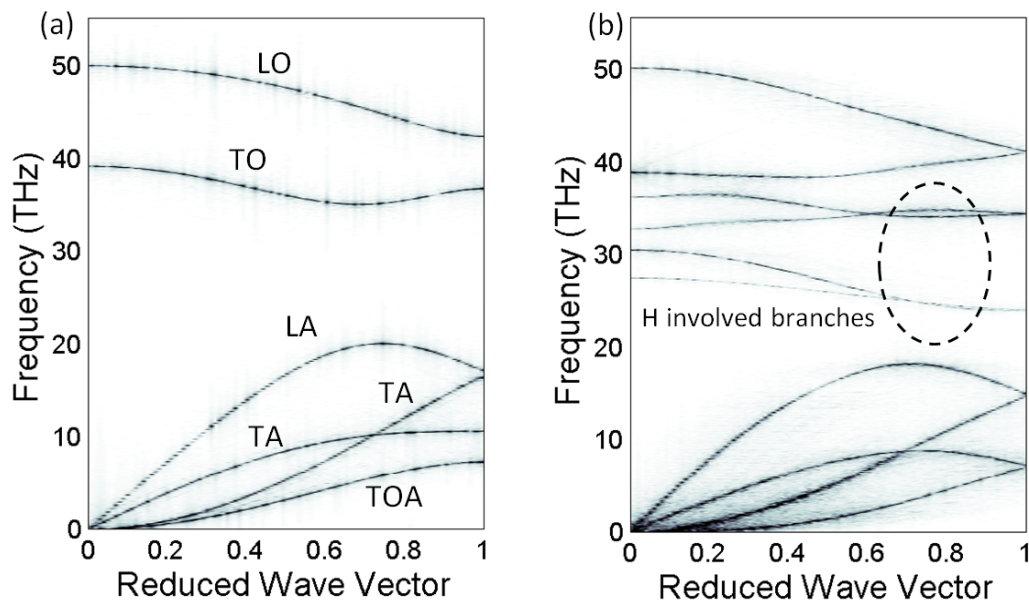

*Figure 4. Phonon dispersion relationship of PA from **(a)** united atom force field and **(b)** COMPASS potential.*

Phonon dispersion calculated from our developed force field shows four acoustic branches: one longitudinal branch (LA), one torsional branch (TOA) and two traverse branches (TA) (Figure 4a). In addition, two optical branches are also observed (Figure 4a, LO & TO). These phonon branches are also observed in COMPASS potentials calculations (Figure 4b). However, some branches in COMPASS potential model meet with each other at the zone boundaries, which is different from the united atom model. The reason is that the partial double bond is used to model the carbon backbone bonds in PA in the COMPASS potential, and this eliminated the difference between C-C and C=C bonds. The same backbone bonds for all carbon/carbon connection impose additional "artificial" symmetry, which causes degeneracy of some vibrational modes at the zone boundary. By distinguishing C-C and C=C backbone bonds, our united atom force field can better capture the chemical nature and vibration mode over COMPASS potentials for PA simulations.

## 4. Conclusions

In this work, DFT energy hypersurfaces are used to parameterize a united-atom force field for PA. Bonding interactions for the alternating single and double bonds in the conjugated polymer backbone are explicitly described and Class II anharmonic functions are separately parameterized. Bond angle and dihedral interactions are also anharmonic and parameterized against DFT energy surfaces. The established force field is then used in MD simulations to calculate the thermal conductivity of single PA chains and PA crystals with different simulation domain lengths. It is found that the thermal conductivity values of both PA single chains and crystals are very high and are length-dependent. At ~790 nm, their respective thermal conductivity values are ~480 W/mK and ~320 W/mK, which are comparatively higher than those of PE – the most thermally conductive polymer fiber measured up-to-date.

**Acknowledgements**

The authors would like to thank the DuPont Young Professor Award and U.S. National Science Foundation (1949910). The computation was supported in part by the University of Notre Dame, Center for Research Computing, and NSF through XSEDE resources provided by TACC Stampede-II under grant number TG-CTS100078.